\documentclass{article}

\usepackage[separate-uncertainty = true]{siunitx}
\usepackage{multirow}
\usepackage{authblk}
\usepackage{graphicx}
\usepackage{geometry}
\usepackage{url}
\usepackage{enumitem}
\usepackage{tabularx}
\usepackage{booktabs}
\usepackage[sorting=none,style=nature]{biblatex}

\begin{document}
\title{Quantitative phase nano-imaging with a laboratory source}

\author[1*]{Luca Fardin}
\author[2]{Chris Armstrong}
\author[1]{Alberto Astolfo}
\author[3]{Sebastian Ignacio Allen Binet}
\author[3]{Matthieu N. Boone}
\author[4]{Rebecca Fitzgarrald}
\author[4]{Yong Ma}
\author[4]{Alexander Thomas}
\author[5]{Darren J. Batey}
\author[1]{Alessandro Olivo}
\author[1*]{Silvia Cipiccia}

\affil[1]{\raggedright\small\itshape{Department of Medical Physics and Biomedical Engineering, University College London, London, WC1E 6BT, UK}}
\affil[2]{\raggedright\small\itshape{Department of Plasma Physics, Central Laser Facility, RAL. OX11 0QZ, United Kingdom}}
\affil[3]{\raggedright\small\itshape{Department of Physics and astronomy, Ghent University, Ghent, Belgium}}
\affil[4]{\raggedright\small\itshape{Gerard Mourou center for ultrafast optical science, University of Michigan, Ann Arbor, Michigan, 48108, USA}}
\affil[5]{\raggedright\small\itshape{Diamond Light Source, Harwell Science and Innovation Campus, Didcot, OX11 0DE, UK}}

\affil[*]{\raggedright\small\itshape{luca.fardin@elettra.eu, s.cipiccia@ucl.ac.uk}}

\date{}
\maketitle

\begin{abstract}
Investigating the structure of matter at the nanoscale non destructively is a key capability enabled by X-ray imaging. One of the most powerful nano-imaging methods is X-ray ptychography, a coherent diffraction imaging technique that has become the go-to method at synchrotron facilities for applications ranging from brain imaging to battery materials. However, the requirements in terms of X-ray beam quality have limited its use to large synchrotron facilities and, to date, only one attempt has been made to translate the technique to a small-scale laboratory. To unleash the power of this technique to the broad user community of laboratory X-ray sources, there are outstanding questions to answer including whether the quantitativeness of the information is preserved in a laboratory despite the drastic decrease in X-ray flux of several orders of magnitude, with respect to synchrotron instruments. In this study not only we demonstrate that the quantitativeness of X-ray ptychography is preserved in a laboratory setting, but we also apply the method to the imaging of a brain tissue phantom. Finally, we describe the current challenges and limitations, and we set the basis for further development and future directions of quantitative nano-imaging with laboratory X-ray sources.
\end{abstract}

\section{Introduction}

Compact X-ray sources suitable for small-scale laboratories have become a common tool for imaging applications in industry, medicine, research and education.
Compared to large facilities such as synchrotrons and X-ray free electron lasers, they are cheaper in building and operational costs, therefore easily available, with benefits in terms of both research and knowledge transfer. However, these advantages come at the cost of reduced quality (coherence) of the X-ray beam and flux, two properties high-resolution imaging techniques often rely on. The high demand for high-quality accessible sources has triggered a scientific effort to overcome these limitations,  both on the imaging technique side and on the source development side. Regarding imaging methods, several advanced techniques developed at large-scale facilities have been successfully translated to small-scale laboratories \cite{Olivo_review_2021,Pfeiffer_gratings_2006,Ronan_lmj_directionaldf,thuring_2013,zanette_2014}. Regarding source development, of note are nanofocus systems and those based on liquid metal jet (LMJ) anode. The former, by using advanced electron optics, have achieved nanoscale spot sizes enabling high resolution imaging. The latter, by replacing a solid anode with a liquid one have reached a brilliance, which is a measure of the coherent flux of the beam, of at least one order of magnitude higher than conventional solid-target microfocus sources, by focusing high electron currents onto micrometric focal spots while avoiding the melting of the anode \cite{Hemberg_lmj_2003,Otendal_2005,tuohimaa}. These innovations enabled a series of breakthroughs. Laboratories based on nano-focus sources took advantage of the small source size and high geometrical magnifications to reach spatial resolutions down to a few hundreds of nanometers, with applications in the study of semiconductors and microelectronics \cite{DREIER_packaging} and virtual histology \cite{Eckerman_VH,Dreier_2024}. Micro-focus sources, coupled with capillary optics, were recently used in a study of laboratory-based transmission X-ray microscopy, with a reported spatial resolution of \qty{100}{\nano\meter} \cite{kukutova_txm}.

The image contrast in the above mentioned applications is mainly based on the X-ray absorption properties of the sample. For low absorbing samples, such as biological specimen, propagation-based phase-contrast imaging has been used \cite{Eckerman_VH,Dreier_2024}. With respect to absorption imaging , this technique uses a different physical process to improve the contrast, namely diffraction in the near-field. Soft tissues, almost invisible to conventional X-ray imaging, can be measured with unprecedented sub-micron spatial resolutions \cite{Eckerman_VH}.
Propagation-based phase-contrast is well adapted to laboratory sources, for its relatively moderate requirements on beam coherence\cite{Eckerman_VH}, but it is in general non-quantitative \cite{paganin_book_2006}. A successful attempt to perform quantitative X-ray phase imaging with a compact X-ray source was published in 2021 by Batey et al., who performed the first far-field ptychographic imaging using a micro-focus LMJ source \cite{batey_ffp_lmj_2021}. X-ray Far-field ptychography (X-FFP) is a lensless coherent diffraction imaging technique \cite{rodenburg_07}. In a standard X-FFP acquisition the sample is scanned at overlapping steps across a spatially confined illumination, usually few micrometer in diameter, while the resulting diffraction patterns are recorded in the far-field without using any imaging lens in front of the detector. The imaging lens is replaced by iterative algorithms \cite{Maiden_2009,Thibault_2009} that combine the diffraction patterns to reconstruct a quantitative map of the absorption and phase induced by the sample on the impinging radiation field as well as the intensity and phase of the spatially confined illumination.
The absence of imaging lenses makes it possible to achieve spatial resolutions limited theoretically only by the diffraction limit: in the X-ray regime at synchrotron facilities, a resolution of \qty{4}{\nano\meter} was recently reported \cite{aikudas_4nm_pty}. 
In the first proof-of-principle of laboratory based X-FFP, Batey et al. achieved a spatial resolution just below \qty{1}{\micro\meter}. Two main factors were identified as limiting the image quality: the coherent flux and the pointing instability of the X-ray beam. For the latter, a phase tilt correction routine within the reconstruction algorithm \cite{batey_ffp_lmj_2021} was developed. The quantitativeness of the retrieved phase was not evaluated in the 2021 study.

In this work, we demonstrate that lab X-FFP can achieve nanoscale resolution better than \qty{300}{\nano\meter}, while providing quantitative 2-D maps of the sample's induced phase shift. We describe in detail the experimental setup allowing the achievement of the results and we show the potential of laboratory based X-FFP through the example of imaging a brain tissue phantom. 


\section{Results}

\begin{figure}
    \centering
    \includegraphics[width=0.8\linewidth]{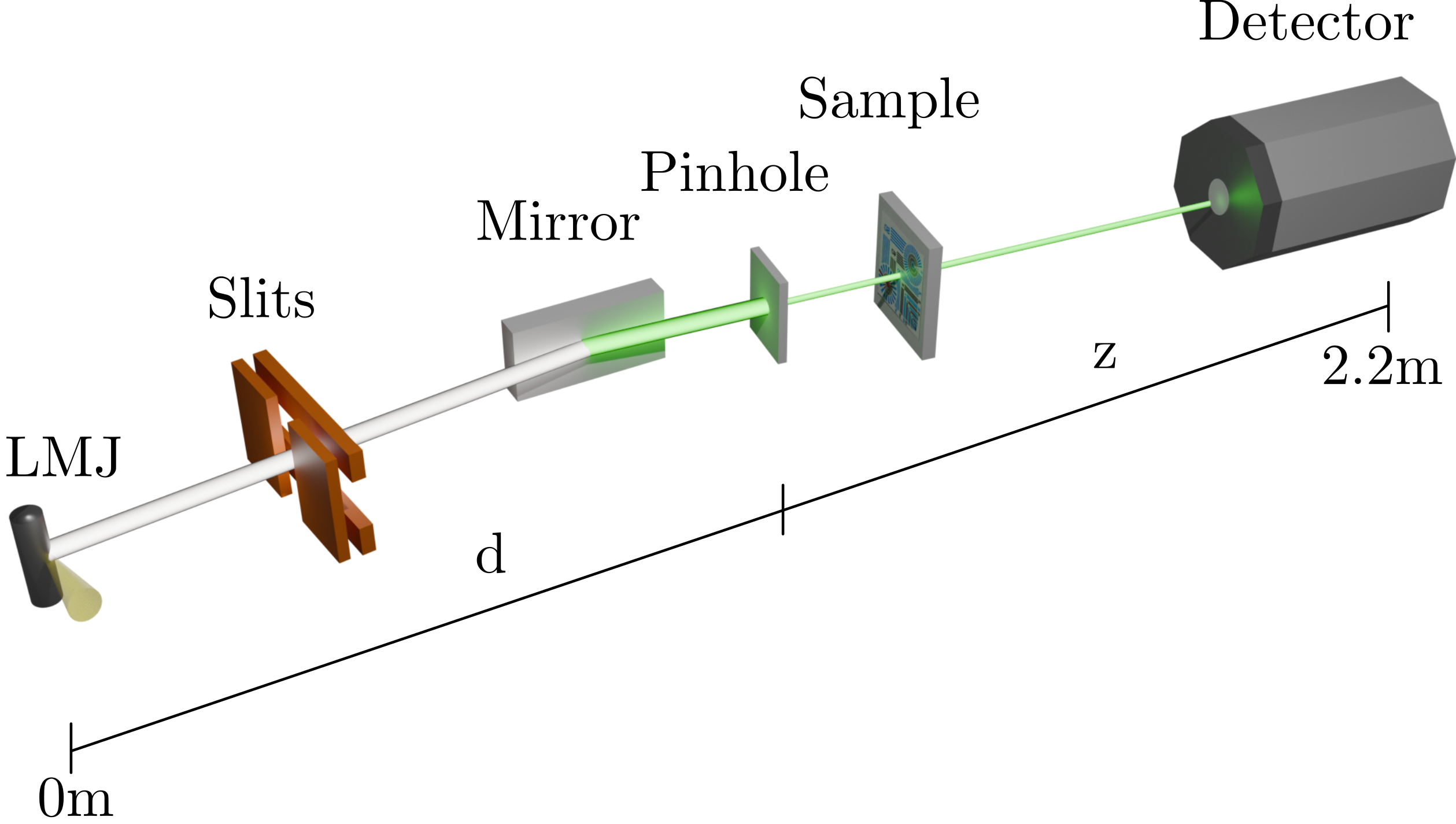}
    \caption{X-FFP setup. The radiation is produced by high-brilliance microfocus LMJ source. A coherent beamlet is isolated by a pinhole and used to scan the sample. Diffraction patterns are detected in the far-field.}
    \label{fig:setup_art}
\end{figure}

The main requirements for an X-FFP setup are brilliance, coherence and stability.
The laboratory-based X-FFP setup used for this study is based on a micro-focus LMJ source with Ga-In alloy. A detailed description of the setup is given in the Methods section and a representation is provided in Figure \ref{fig:setup_art}. In the absence of focusing optics, the lateral coherence length of the radiation $\epsilon_l$ at a distance d from the source can be estimated as $\epsilon_{x,y} \approx \lambda d /\ s_{x,y}$ \cite{wolf2007introduction}, where $\lambda$ is the wavelength and $s_{x,y}$ the lateral dimension of the source. The lateral coherence of the beam can be increased by reducing $s_{x,y}$, which in the case of a LMJ source is determined by the projected focal spot size of the electron beam on the metal jet. In our system, the focal spot size could be tuned down to a minimum of \qtyproduct{15 x 15}{\micro\meter}. As the focal spot size decreases, the maximum power is automatically limited, to limit the power load on the metal-jet and avoid the formation of metal vapors.  A pinhole with diameter D matching $\epsilon_l$ is used to select the coherent fraction of the beam, thus creating an ideally coherent beamlet of size $\approx D$ illuminating the sample. Unlike reference \cite{batey_ffp_lmj_2021}, a pinhole has been preferred to a focusing optics to maximize the stability of the setup, which is therefore mainly affected by fluctuations of the source and external vibrations.
The sample-to-detector distance  $z$ 
is chosen in X-FFP to satisfy the far-field condition at the detector $D^2/\ (\lambda z)<<1$ \cite{paganin_book_2006} and to ensure sufficient sampling in the frequency domain $D\approx \lambda z /\ 2\Delta x$ \cite{SPENCE}, where $\Delta x$ is the pixel size of the detector.
Undersampling in the frequency domain can be dealt with in X-FFP, thus relaxing the second condition, if the step-size of the scan is reduced, at the cost of an increased number of scanning points for a given field-of-view \cite{batey_th_2014}, a critical parameter when using low-brilliance sources.

The temporal coherence length is defined by using a flat multi-layer monochromator to select the Ga $K_{\alpha}$ emission line at \qty{9.25}{\kilo\electronvolt}.

Based on the above considerations, a \qty{5}{\micro\meter} pinhole was positioned at \qty{1}{\meter} from the source, where the spatial coherence length is expected to be approximately \qty{8}{\micro\meter}, for a selected source size of \qty{20}{\micro\meter}. The sample was placed just downstream of the pinhole and the detector, featuring a pixel size of \qty{13}{\micro\meter}, was placed at \qty{1.2}{\meter} from the sample, providing sufficient sampling in the Fourier space and satisfying the far-field condition with $D^2 /\ \lambda z = 0.2$. As a side note, the size and location of the pinhole are constrained in the described setup by the maximum length of our compact setup, which is equal to \qty{2.5}{\meter}. 

\subsection{Setup characterization}

\begin{figure}[htbp]
\centering\includegraphics[scale=0.3]{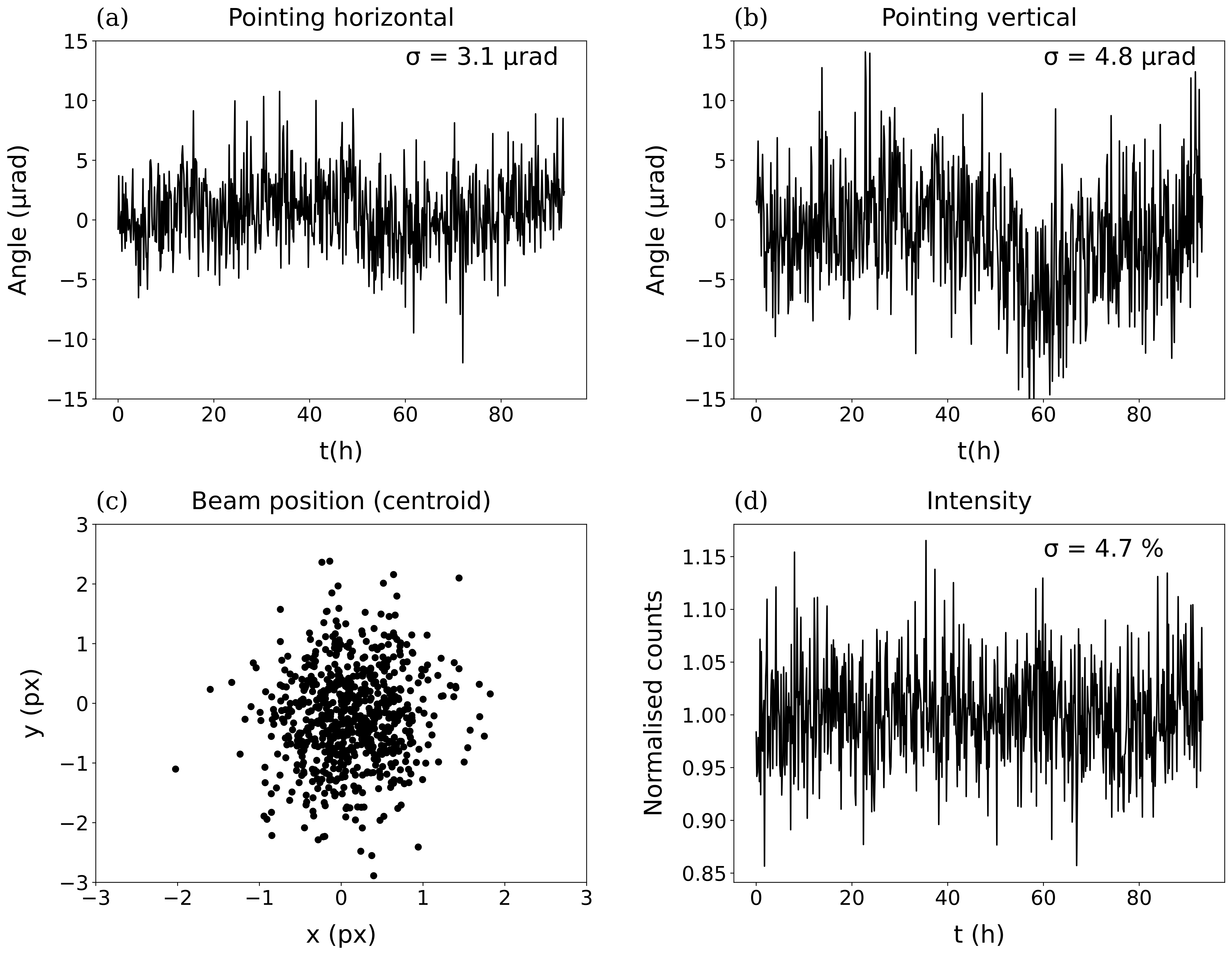}
\caption{Stability of the X-ray source over just less than four days. Horizontal (a) and vertical (b) pointing stability and associated standard deviations, measured from the time evolution of the centroid of the beam. (c) Scatter plot of the measured beam centroids, with respect to the centroid of the median beam, as recorded by the detector. (d) Integrated beam intensity, as a function of time, normalized by the median and associated standard deviation.}
\label{fig:stability}
\end{figure}

As the beam stability and coherent flux were the main limitations found in the previous study on laboratory-based X-ray ptychography \cite{batey_ffp_lmj_2021}, the setup used for the experiment in this work has been firstly characterised by looking at these two properties.
The stability has been  measured by tracking the image of the pinhole on the detector for \qty{90}{\hour}.
The results are shown in Figure~\ref{fig:stability} (a-d). The centroid of the beam fluctuates mostly within one pixel, with peaks up to 3 pixels in the vertical direction (Figure~\ref{fig:stability} (c)). The standard deviation is of 0.5 pixels in the horizontal direction and 0.8 pixels in the vertical direction, corresponding to \qty{3.1}{\micro\radian} and \qty{4.8}{\micro\radian} respectively. The time evolution of the fluctuations (Figure ~\ref{fig:stability} a-b) shows a low frequency trend, mostly confined within 1 pixel (\qty{6}{\micro\radian}), with one sharp displacement of 1.5 pixel occurring in the vertical direction after \qty{60}{\hour}.
These fluctuations are most probably associated with instabilities in the electron beam and thermal instabilities of the setup, with consequent movements of the pinhole with respect to the detector. Although the temperature in the laboratory is kept constant within a degree via a standard air conditioning unit, the source and the setup are enclosed in two connected cabinets for radiation shielding. The cabinets are not temperature controlled, but only equipped with extraction fans.
The intensity remained constant throughout the measurement, with variations around the median value below \qty{5}{\percent} (root mean squared error). Its time evolution and standard deviation are shown in Figure~\ref{fig:stability} (d).

\begin{figure}[htbp]
\centering\includegraphics[scale=0.3]{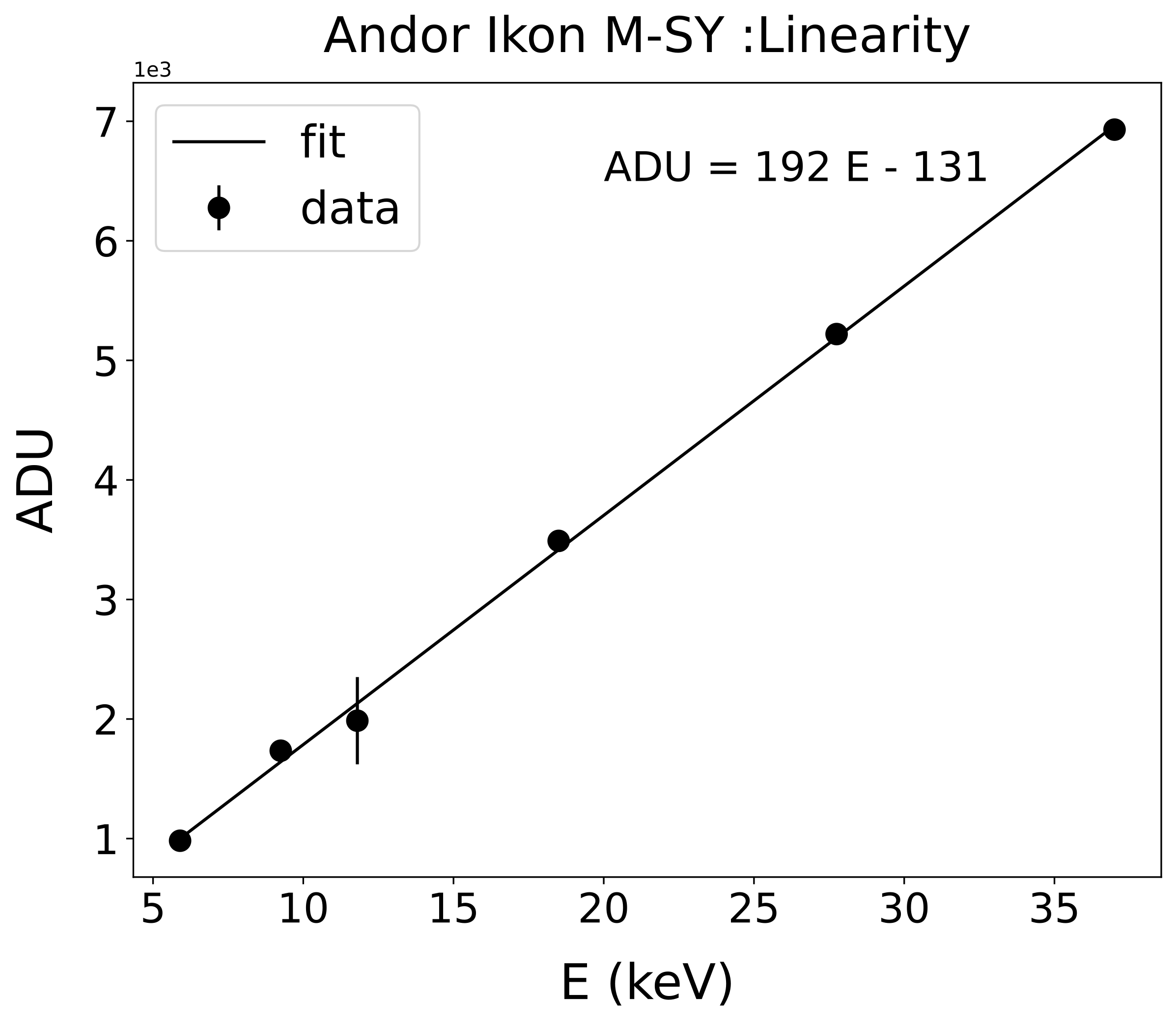}
\caption{Linearity of the Andor Ikon M-SY CCD detector. See the Characterization section in the Methods for details on the protocol used.}
\label{fig:linearity}
\end{figure}

The X-ray flux impinging on the detector was evaluated by first assessing the linearity of the detector, an Andor Ikon M-SY, as detailed in the Methods section.
The detector response shows a good linearity (Figure \ref{fig:linearity}) and we have estimated a conversion factor of 190 analog digital units (ADU) per \qty{}{\kilo\electronvolt} of energy absorbed, with an intercept of -130 at gain 4. This relation allows for an approximate conversion of the signal integrated during the intensity stability to a number of photons impinging over the integration time. Once corrected for the efficiency of the detector ($\approx 30\% $ at \qty{9}{\kilo\electronvolt}), this relation provided a measurement of the flux at the detector of approximately 5 photons per second. The low flux imposes a long integration time per diffraction pattern, set in this study to 8 minutes. 

The size of the source was measured to confirm the nominal value and was estimated from the source-induced blur on the image of a \qty{20}{\micro\meter} pinhole (pinhole-camera approach, see Methods section). The requested source size was set as $s_{x,y} = $ \qty{20}{\micro\meter} and the experimentally measured size was found to be \qty{26\pm2}{\micro\meter} ($s_x$) and \qty{15\pm2}{\micro\meter} ($s_y$). The accuracy of the electron spot-size on the LMJ can be affected by the aging of the cathode and the quality of the electron-beam alignment.


\subsection{Quantitative phase imaging}

\begin{figure}[htbp]
\centering\includegraphics[scale=0.35]{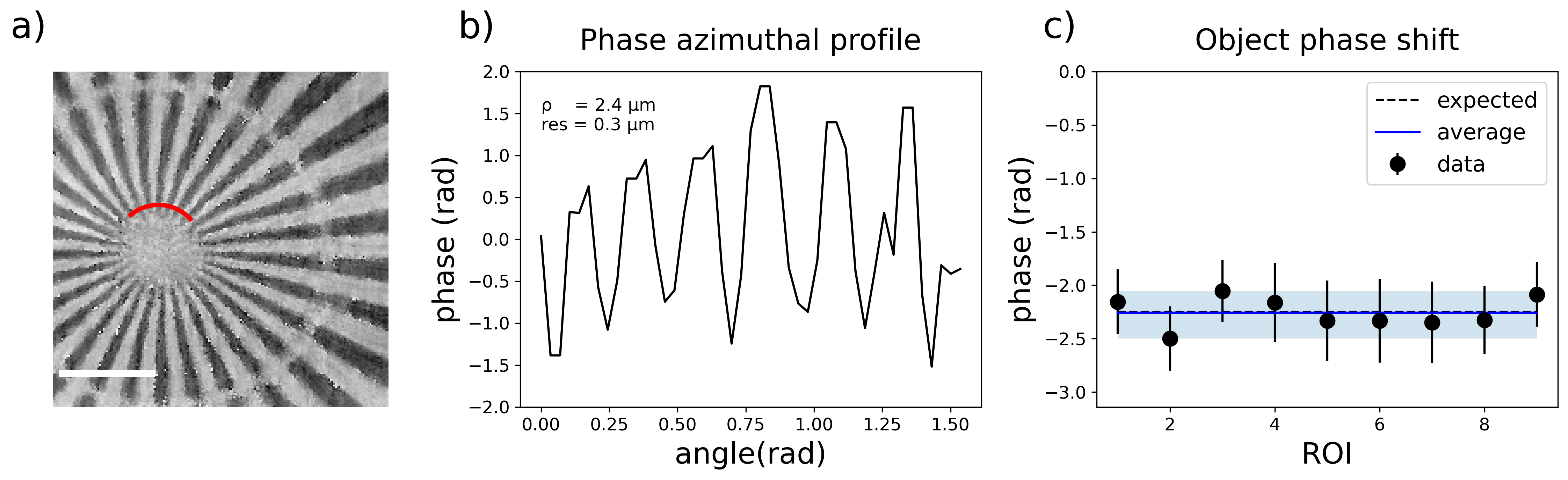}
\caption{Quantitative phase reconstruction of a Siemens star test sample. (a) Phase image reconstructed by 20000 iterations of the extended ptychographic engine algorithm (ePIE) \cite{Maiden_2009}, The scale bar corresponds to \qty{5}{\micro\meter}. (b) average phase shift of the star (Au) compared to the surrounding protective polymer. It was calculated as the relative phase of ten couples of ROIs. The mean value, with uncertainty band is shown in blue and compared to the expected value. (c) azimuthal profile of the phase corresponding to the red line in (a). The radial distance $\rho$, as well as the minimum spoke width (res, i.e. resolved feature), is reported.}
\label{fig:quantitative}
\end{figure}

The quality of X-ray ptychography achievable with our setup, was tested with a Siemens star resolution target, consisting of \qty{1.5}{\micro\meter} thick Au spokes with a minimum width of \qty{200}{\nano\meter}. The results are shown in Figure~\ref{fig:quantitative}, which corresponds to a \qtyproduct{20x20}{\micro\meter} field of view acquired in 48 hours.
The retrieved phase and a quantitative analysis are shown in Figure~\ref{fig:quantitative} (a) and (b) respectively. The measurement of the phase shift induced by the star, with respect to the surrounding polymer layer, was obtained by comparing 9 pairs of regions of interest (ROI), each pair consisting of ROIs taken inside and immediately outside the stars' spokes. These ROIs had different sizes, to adapt to the variable width of the spokes. This averaged based approach in the analysis was chosen to deal with a low frequency modulation in the background, most probably related to long term instabilities, and with the artifacts at the margin of the image caused by the reduced scan overlap at the edges. The phase shift measured in the sample on the Au with respect to adjacent area without the Au for the 10 ROIs is shown in (b), together with their mean and standard deviation $\sigma_\phi$ (blue line and band, corresponding to $3\sigma_\phi$), and is equal to \qty{-2.255 \pm 0.046}{\radian}.  Their average well matches the expected  value of \qty{-2.25}{\radian}. The expected value was calculated as the relative phase induced by Au and the polymer layer. The refractive index decrements of Au and polymer were estimated using a public online database \cite{Cxro,Cxro_Henke} and information from the maker of the Siemens star\cite{xrnanotech}, respectively.   
The spatial resolution of the scan was estimated from an azimuthal profile across the spokes, close to the center of the star (red line in (a)), and was defined as the minimum size of the spoke that is visible in the reconstruction. This value, estimated as half of the curve period in (c), was found to be \qty{300}{\nano\meter}.

\subsection{Brain Imaging}

\begin{figure}[htbp]
\centering\includegraphics[scale=0.62]{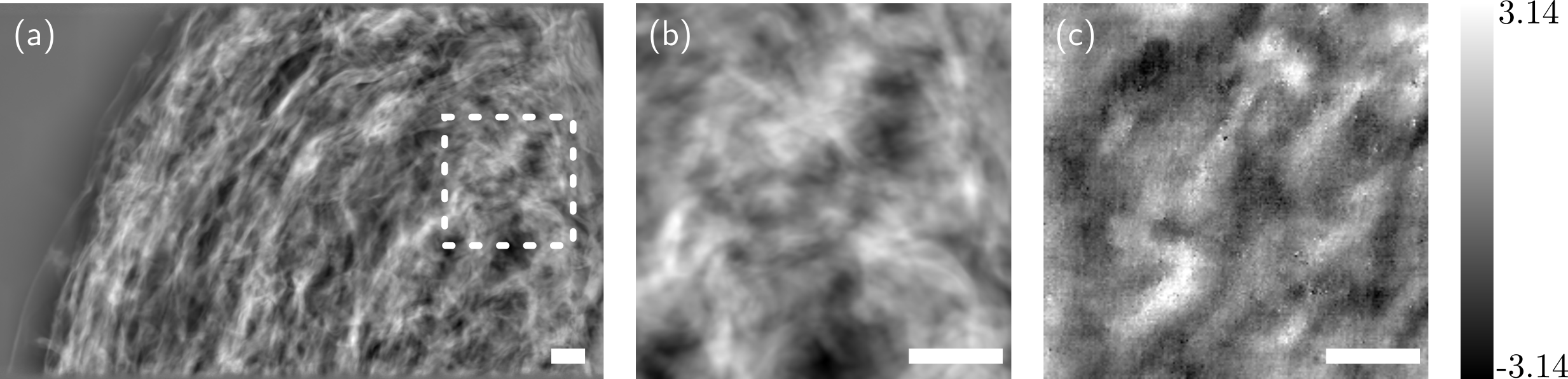}
\caption{Ptychographic scan of a polymer used as brain phantom for diffusion magnetic resonance imaging. (a) High resolution scan acquired at the Diamond Light Source, adapted from \cite{Erin_2025}. (b) Detail from (a). (c) Lower resolution scan acquired with the laboratory-based ptychography setup. (b) and (c) have the same spatial extent but do not correspond to the same region. The scale bar corresponds to \qty{5}{\micro\meter} in all images.}
\label{fig:brain}
\end{figure}

A mainstream application of ptychography is nanoimaging of low-absorbing materials, such as biological tissues\cite{cipiccia_fast_ptycho, bosch_2025}. To demonstrate the capabilities of laboratory based X-FFP beyond a test sample, we imaged on a more heterogeneous sample of interest for biology, namely a polymer that mimicks the brain tissue. This consists of of polycaprolactone microcylinders reproducing the structure of axons. This type of phantoms is used in validation studies for diffusion magnetic resonance brain imaging \cite{Zhou_brain_phantom,McHugh_brain}. Figure \ref{fig:brain} shows the phase reconstruction of the brain phantom (b) compared to one retrieved from a synchrotron scan (a) during a previous study \cite{cipiccia_fast_ptycho,Erin_2025}. The two scans covered different areas of the sample thus making a direct comparison impossible. However, the similarity of the structures and range of the sample induced phase shifts can be appreciated from Figure \ref{fig:brain} (b) and (c). 

\section{Discussion}

This study presents the first compact laboratory setup dedicated to X-FFP and operating at the \qty{9.25}{\kilo\electronvolt} Ga $K_\alpha$ emission line. Unlike a previous study which only showed feasibility \cite{batey_ffp_lmj_2021}, here we demonstrate quantitativeness in the phase reconstruction, and the ability to resolve features of a few hundreds of nanometers. The image quality achievable with this compact setup was demonstrated in a resolution target phantom and in a sample of interest for biomedicine, specifically neuroimaging, one of the many fields that may benefit from this technology.
The scanning time for the results presented in this work, and corresponding to a \qtyproduct{20x20}{\micro\meter} field of view, was approximately 50 hours, driven mainly by the 8 minutes integration time of each diffraction pattern. 
To unsure the suitability for long acquisition times, the setup was characterized for long-term stability, over a 90-hour time interval. 
The results showed beam spatial fluctuations on the detector occurring mainly within one pixel and intensity fluctuations of $5\%$, comparable in amplitude to those observed in the 2021 study (Figure 3 in \cite{batey_th_2014}). The main difference appears to be the x-ray beam intensity time evolution: while fast fluctuations of the order of minutes were reported in \cite{batey_th_2014}, we measured only long term fluctuation of the order of hours (see Figure \ref{fig:stability} (d)). 

A second effect of long integration times is the noise due to cosmic rays interacting with the detector. Lead shielding was added around the camera to reduce the number of interactions and a dedicated post-processing routine has been developed to remove the longest traces generated by the cosmic rays (see Methods section).
Two additional sources of image noise are electronic noise and air scattering. To reduce the camera's electronic noise, air cooling was set at \qty{-70}{\degreeCelsius}.This could be further improved by using coolant chillers down to \qty{-100}{\degreeCelsius}, which were not available in this experiment. On-axis scattering from the air has been minimized in the setup described by installing an \qty{80}{\centi\meter} long vacuum pipe between the sample and the detector. Reducing scattering from the air is especially critical, since it produces a signal with a similar statistics to that from the scattering from the sample, especially far from the center of the beam, making it difficult to discriminate between signal and noise. 

The ptychographic imaging produced in this study shows the first quantitative phase reconstruction obtained with a laboratory source. Moreover, with respect to the first proof of concept \cite{batey_ffp_lmj_2021} the spatial resolution is enhanced by well over a factor three and a value of approximately \qty{300}{\nano\meter} is measured. The main factors we identify as key for the achieved results are the stability and simplicity of the setup. While in ref \cite{batey_ffp_lmj_2021} scan positions had to be dropped due to drastic changes in intensity and a phase tilt correction was made necessary by the pointing fluctuations, in this work only position correction \cite{MAIDEN201264} was used to deal with the long term drift of the setup. The higher stability could be due to the source (a newer model, the D2+, is used for this experiment, while the D2 in ref \cite{batey_ffp_lmj_2021}) as well as to different pre-sample optics. The multilayer focusing optics used in the 2021 study was replaced by a flat mirror in this work, which could be more resilient to thermal variation. However, these are only hypotheses as in the current study it was not possible to decouple the effect of the source from that of the optics. Moreover, the use of a pinhole to define the illumination at the sample has removed any uncertainties on the beam coherence properties at the focus and on its position unlike in reference \cite{batey_ffp_lmj_2021}. However, it should also be noted that the sample used in ref \cite{batey_ffp_lmj_2021} produced a weaker phase shift of 0.8, which is likely to have contributed to making quantitative phase determination more difficult. 

Future work on laboratory X-FFT should focus on reducing the acquisition time of the scan, which is currently subject to long term instabilities, affecting sensitivity and resolution, and preventing 3D applications. An efficient way to access larger fields of view could be obtained efficiently by illuminating the sample with multiple mutually incoherent probes \cite{Lyub_multibeam}. This can be achieved by using an array of pinholes, separated by a distance larger than the coherence length, thus utilizing a larger fraction of the beam flux. This technique, called multi-beam ptychography, has been proven to decrease the scanning time for a given field of view compared to conventional single-beam ptychography while preserving the image quality \cite{Lyub_multibeam}. A different approach, called single-shot ptychography \cite{Sidorenko_singleshot}, would eliminate the need for scanning, thus reducing the acquisition time from 50 h to that of a single step exposure, in hour case 8 minutes. This approach relies on multiple beamlets being focused on the sample, with an inter-beamlet spacing corresponding to the conventional scanning step. Each beamlet produces a diffraction pattern, that is individually recorded on the camera. This method, already tested in the soft and tender X-ray regime \cite{kharitonov_singleshot,Levitan_singleshot}, requires the use of a mirror and a mask to generate and focus the beamlets on the sample.

Simpler and more immediate solutions include increasing the source size to enable a higher electron current and therefore X-ray flux, at the cost of lateral coherence. A modest decrease in lateral coherence could be compensated by increasing the number of orthogonal probe modes in the reconstruction \cite{thibault_state_mix_2013}.
The low photon statistics could be partially mitigated by a more efficient detector. According to the manufacturer, the Andor Ikon M-SY has an efficiency of $30\%$ at \qty{9}{\kilo\electronvolt}. Single-photon-counting detectors have efficiency close to 100\% at \qty{9}{\kilo\electronvolt}, but a larger pixel size (usually in excess of \qty{50}{\micro\meter}), that requires longer sample-detector distances to satisfy the sampling condition in the frequency domain.
While in our compact setup their use is hindered by the limited length of the radiation shielded area, this could be easily overcome with few extra meters of propagation in future system designs.

Summarising, a laboratory X-FFP has the potential to become a wide spread, easily accessible tool for nanoimaging. There are several possible approaches to overcome its current limitations, most of which can be implemented with relatively small investments in space and equipment. We also foresee that by virtue of its simplicity, the setup will be compatible with new emerging compact high-brilliance X-ray sources, such as those based on laser-plasma interaction \cite{ELI_2024} or inverse-Compton scattering \cite{VanElk:25}.

\section{Methods}

\subsection{Experimental setup}
The experiments were carried out at the CLF-AXIm  Scientific Collaboration (CASC) laboratory at the Rutherford Appleton Laboratory in Oxfordshire, UK. 
The laboratory is equipped with a D2+ liquid-metal jet X-ray source, produced by Excillum (Excillum AB, Sweden). The source can operate at a maximum voltage of \qty{160}{\kilo\volt} and a maximum electron beam power of \qty{250}{\watt}. In this study it was run at \qty{70}{\kilo\electronvolt} and  \qty{1}{\milli\ampere} \qty{70}{\watt}. The maximum power is automatically adjusted based on the source size, to limit the power load on the metal-jet and to avoid the formation of metal vapors.
The liquid anode is the I1 alloy, made of Indium, Gallium and Tin.  

A scheme of the experimental setup is shown in Figure \ref{fig:setup_art}. The X-ray beam is shaped by vertical and horizontal slits placed approximately 40 cm downstream of the source to minimise scattering from the beam impinging on the lead radiation shielding around the experimental area. The beam is then monochromatized by using a single multilayer mirror (AXO Dresden GmbH, Dresden, Germany). The $k_{\alpha}$ emission line of Gallium (\qty{9.25}{\kilo\electronvolt}) is selected, to maximize the available monochromatic flux. The deflection angle is 2.4\textdegree. 

The sample illumination is defined with a \qty{5}{\micro\meter} pinhole, laser micomachined on a \qty{50}{\micro\meter} thick tungsten foil (Scitech Precision, Didcot, UK). The pinhole and the sample are both placed \qty{1}{\meter} from the source and are both mounted on two high-precision 3-axis translation stages (XYZ-SLC23:30, SmarAct GmbH, Germany), for positioning and scanning. 

The detector is \qty{1.2}{\meter} downstream of the pinhole-sample assembly. This is the maximum distance available at CASC, due to experimental constraints: the setup is fully contained within a \qty{2}{\meter} long lead shielded cabinet (EZ-Access X-Ray Cabinet, Metrix NDT, Laughborough, UK), that ensures radiation protection while limiting the maximum beam propagation distance. An \qty{80}{\centi\meter} vacuum tube was placed between the sample and the detector, to reduce scattering and absorption from air, that at 9.25 keV for 80 cm of air at 1 bar is about \qty{35}{\percent}. 

The detector is an Andor Ikon M-SY (Oxford Instruments, Abingdon, UK). It features a back-illuminated CCD sensor, protected by a \qty{200}{\micro\meter} beryllium window. It has a \numproduct{1024 x 1024} sensor array with \qtyproduct{13 x 13}{\micro\meter} pixels and has a nominal quantum efficiency of \qty{30}{\percent} at \qty{9}{\kilo\electronvolt}. It was operated in a temperature of \qtyrange{-60}{-70}{\degreeCelsius}, to minimize the dark noise, with a preamplifier gain of 4. A further 5 mm thick lead sheet was positioned around the detector to mitigate the effects of background and cosmic radiation.

\subsection{Characterization}

The setup was characterized for pointing and intensity stability, linearity and source size. The measurement of pointing stability was carried out by acquiring 2800 consecutive images of the \qty{5}{\micro\meter} pinhole over approximately 90 hours, each with an exposure time of 2 minutes. To improve the photon statistics of a single projection, 4 consecutive images were added together, following the procedure described in Methods section \ref{sec:im_proc}, providing an effective number of 700 acquisition points with 8 minutes of exposure. The pointing stability was measured by determining the position of the center of mass of the beam on the detector as a function of time. Movements were calculated against the center of mass of the median of all projections. The intensity stability was retrieved from the same dataset, by computing the integrated flux on the detector as a function of time.

The linearity of the detector was measured by exposing the camera to the X-ray emission of a $^{55}\mathrm{Fe}$ source, with a known energy of \qty{5.9}{\kilo\electronvolt}. The exposure time was set to \qty{0.5}{\second}. A gaussian fit was performed on the two peaks of the spectrum of analog digital units (ADU), corresponding to single and double photon interactions. The same measurement was repeated with the $K_\alpha$ line of gallium, selected from the LMJ spectrum with the multilayer monochromator, which providedd spectral points using from single up to quadruple counting. Linearity was verified by a linear fit of the ADUs as a function of the total energy absorbed in the pixels.

The source size was determined with a pinhole camera approach, by using an array of 18 pinholes of \qty{20}{\micro\meter} micromachined on a \qty{50}{\micro\meter} thick tungsten foil, equally spaced at \qty{60}{\micro\meter}, and placed at \qty{1}{\meter} from the source. The \qty{5}{\micro\meter} pinhole used in ptychography scans was removed for this measurement. If the image of each pinhole, on the detector, is assumed to have a Gaussian intensity distribution, its variance $\sigma^2_{Im}$ is given by the sum of the magnified pinhole diameter $\sigma_{Pin} =D_{Pin}/\sqrt{12}$ and the magnified source size according to:
\begin{equation}
\label{eq:source}
    \sigma^2_{Im} = \left( \frac{z+d}{d} \sigma_{Pin} \right)^2 + \left( s_{x,y} \frac{z}{d} \right)^2
\end{equation}
where $d$ and $z$ are the source-to-pinhole and the pinhole-to-detector distance respectively. The contribution of diffraction is negligible in this geometry. $(z+d) /\ z$ is the magnification of the system, which can be estimated as the ratio between the measured and the nominal center-to-center distance between the pinholes.
To estimate the magnification, an ideal grid that defines the expected unmagnified distribution of the centroids of the pinholes was registered onto the measured magnified centroids, by applying an affine image registration model. The source size $s_{x,y}$ was determined by inverting Eq. \ref{eq:source}. The minimum source size that can be measured in our geometry is \qty{10}{\micro\meter}.

\subsection{Ptychography}

For the ptychographic scan, a source size of \qtyproduct{20 x 20}{\micro\meter} was selected, corresponding to an estimated coherence length at the pinhole position of \qty{8}{\micro\meter}. The high voltage was set to \qty{70}{\kilo\electronvolt} and the power limited to \qty{70}{\watt}. The scan was performed over a \qtyproduct{20 x 20}{} regular grid, with a step size of \qty{1}{\micro\meter}. At each position, 4 independent measurements of the diffraction pattern were performed, with an exposure time of \qty{120}{\second} each. These projections were treated according to the procedure described in Section \ref{sec:im_proc}.
The scans were reconstructed using the ePIE algorithm \cite{Maiden_2009} implemented in the PtyREX code \cite{batey_th_2014}. Projections were cropped to a \qtyproduct{128 x 128}{} window, providing a reconstructed pixel size of \qty{96.6}{\nano\meter} in the object plane. 

\subsection{Samples}
Ptychography was tested on two samples: a Siemens star test pattern (XRnanotech GmbH, Villigen, Switzerland) and a biomimetic brain tissue phantom \cite{Zhou_brain_phantom}. The Siemens star consists of \qty{1.5}{\micro\meter} thick golden spokes on a silicon nitride substrate that are surrounded by a protective polymer. The spoke width decreases from \qty{500}{\micro\meter} on the outermost edge of the star to \qty{200}{\nano\meter}, close to the center.
The biomimetic brain tissue sample is a polymer made of polycaprolactone hollow microfibers, that mimics the axons in the white matter. Its structure and density are close to that of brain tissue and was developed to be used in validation studies of diffusion magnetic resonance brain imaging \cite{McHugh_brain}. This sample was extensively characterized
 with X-rays in a previous study \cite{cipiccia_fast_ptycho}. The phantom was cut with a scalpel to a thickness of approximately \qty{100}{\micro\meter} and imaged with the same protocol used for the Siemens star and described in the previous section.

\subsection{Image processing}\label{sec:im_proc}
All images acquired during this project were collected as a series of 4 consecutive exposures of 2 minutes. The 4 consecutive projections were summed after an individual background removal. The sum was preferred to the median as, due to the low flux of the source, the use of the median heavily suppressed the signal at high spatial frequency, reducing the spatial resolution of the scan. Background images were acquired for this purpose at the end of each scan. Residual time dependent variations in the background of the images required a second image-dependent background estimation, obtained by applying a median filter to each sum with a kernel width of 30 pixels. Cosmic rays traces were afterwards identified as connected regions with a large elongation, estimated as the ratio of the eigenvalues of their inertia tensor. Pixels identified as cosmic rays traces were replaced with the median of the surrounding pixels in a \qtyproduct{5 x 5}{} region. 


\section{Acknowledgments}
This work is supported by EPSRC New Investigator Award EP/X020657/1 and Royal Society grants RGS/R1/231027 and IES/R2/232207. S.I.A.B and M.B. are supported by the FWO G050724N. A.O. is supported by the Royal Academy of Engineering under the Chairs in Emerging Technologies scheme (grant CiET1819/2/78). We would like to thank Tim Arden from I13-1 at Diamond Light Source for his help in the preparation of the vacuum pipe for the CASC lab, and the Thomas Zinn and Paul Wardy from LabSAXS instrument at the Diamond Light Source for their help and suggestions to handle the issues with the LMJ.

The authors thank Michela Fratini, Geoff Parker, Marco Palombo, and Fenglei Zhou for providing the brain-mimicking phantom.

\printbibliography
\end{document}